\documentclass[prl,preprint]{revtex4}
\usepackage{dcolumn,amsmath}
\usepackage{bm}
\newcommand{\tref}[1]{Table~\ref{#1}}

\begin{document}
%############################################################
%\large
\title{In search of the electron dipole moment: {\it Ab initio} calculations
on $^{207}$PbO excited states}.
%on low-lying excited states of $^{207}$PbO}.
\author{T.A. Isaev} \email{timisaev@pnpi.spb.ru}
\author{A.N. Petrov}
\author{N.S. Mosyagin}
%\author{M.G. Kozlov}
\author{A.V. Titov}
\affiliation{Petersburg Nuclear Physics Institute, Gatchina, 188300, Russia}
\author{E. Eliav}
%\email{ephraim@jade.tau.ac.il}
\author{U. Kaldor}
\affiliation{School of Chemistry, Tel-Aviv University, Tel-Aviv 69978, Israel}

\begin{abstract}
We report {\it ab initio} correlated relativistic calculations
of the effective electric field $W_d$ acting on the electron in two
excited electronic states of PbO, required for extracting the
electric dipole moment of the electron from an
ongoing experiment at Yale, which has the potential of improving accuracy
for this elusive property by several orders of magnitude. The generalized
relativistic
effective core potential and relativistic coupled cluster methods are used,
followed by nonvariational one-center restoration of the four-component
wavefunction in the heavy atom core.  $W_d$ is
 $-3.2\times 10^{24}{\rm Hz}/(e\cdot {\rm cm})$ for the $a(1)$ state and
 $-9.7\times 10^{24}{\rm Hz}/(e\cdot {\rm cm})$ for the $B(1)$ state.
Comparison of calculated and experimental values of the hyperfine
constant $A_{\parallel}$ provides an accuracy check for the calculation.
\end{abstract}

\maketitle

\paragraph*{Introduction.}

Following the discovery of the combined CP-parity violation in $K_0$-meson
decay \cite{Christenson:64}, the search for the electric dipole moment (EDM)
of the electron, $d_e$, has become one of the most fundamental problems in
physics \cite{Landau:57}.  Considerable experimental effort has been invested
in measuring atomic EDMs induced by the electron EDM. The best available
results for the electron EDM were obtained in the atomic Tl experiment
\cite{Regan:02}, which established an upper limit of $|d_e|<1.6\cdot10^{-27}$
e$\cdot$cm. It is expected that diatomic molecules containing a heavy atom can
yield more definite results.  Modern experiments searching for $d_e$ in these
molecules exploit the fact that the
%tav: it's actually an electromagnetic field; Timur should know it:
%
% electric field seen by an unpaired electron, $W_d$, is greatly enhanced
% relative to the external field \cite{Hinds:97,DeMille:00}.
%
 effective electric field seen by an unpaired electron, $W_d$, is greatly
 enhanced by the relativistic effects relative to the external field
 \cite{Hinds:97,DeMille:00} reachable in a laboratory.
%tav end
The value of $W_d$ is necessary to extract $d_e$ from experimental
measurements.  For diatomic molecules with one unpaired electron, such as YbF
and BaF, semiempirical estimates or {\it ab initio} calculations with
approximate accounting for correlation and relativity provide reasonably
reliable $W_d$ values (see Refs.\ \cite{Titov:96b,Kozlov:97,Mosyagin:98}).
These molecules are, however, chemical radicals, posing experimental problems.
It was pointed out recently that the excited $a(1)$ \cite{DeMille:00} or
$B(1)$ \cite{Egorov:01} states of PbO can be used effectively in the search
for $d_e$. A novel experiment, using a vapor cell to study excited PbO, has
been started at Yale University.  The unique suitability of PbO for searching
the elusive $d_e$ is demonstrated by the very high statistical sensitivity of
the Yale experiment to the electron EDM, allowing detection of $d_e$ of order
$10^{-31}$ e$\cdot$cm \cite{DeMille:00}, four orders of magnitude lower than
the current limit quoted above. While semiempirical calculations
\cite{Kozlov:02} may be valuable, the authors of \cite{Kozlov:02} stressed
that ``more elaborate calculations were highly desirable''.  High-order {\it
ab initio} correlated relativistic calculations of the type developed recently
\cite{Petrov:02} are required to give accurate values of $W_d$ acting on the
unpaired PbO electrons. An accuracy check is provided by calculating
experimentally known properties which also depend on the electron spin density
near the heavy nucleus, such as hyperfine constants.
% Since the relevant operator is concentrated near the nucleus of the heavy
% atom, the concurrent calculation of experimentally known properties that
% also depend on the electronic spin density near the heavy nucleus, such as
% hyperfine constants, gives a check on the accuracy and reliability of the
% calculated $W_d$.

The terms of interest for PbO in the effective spin-rotational Hamiltonian may
be written following Ref.\ \cite{Kozlov:95}.  The P,T-odd interaction of $d_e$
with $W_d$ is
\begin{equation}
\label{eq:hd}
   H_{\rm edm} = W_d~d_e (\bm{J} \cdot \bm{n}),
\end{equation}
 where $\bm{J}$ is the total electron moment and $\bm{n}$ is the unit vector
 along the axis from Pb to O.  The hyperfine interaction of the
 electrons with the $^{207}$Pb nucleus is
\begin{equation}
\label{eq:hhf}
   H_{\rm hfs}= \bm{J} \cdot \hat A \cdot \bm{I},
\end{equation}
 where $\hat A$ is the hyperfine tensor, characterized for a linear molecule
 by the constants $A_{\parallel}$ and $A_{\perp}$, and $\bm{I}$ is the spin of the
 $^{207}$Pb nucleus ($I=1/2$).

In practice, the effective operator
\begin{eqnarray}
    H_d=2d_e
    \left(\begin{array}{cc}
    0 & 0 \\
    0 & \bm{\sigma E} \\
    \end{array}\right)
\label{Wd}
\end{eqnarray}
 is used to express the interaction of $d_e$ with the inner molecular electric
 field $\bm{E}$ ($\bm{\sigma}$ are the Pauli matrices), to avoid the large
 terms which cancel each other \cite{Martensson:92} because of Schiff's
 theorem.  After averaging over the electronic coordinates in the molecular
 wavefunction, one obtains
\begin{equation}
   W_d = \frac{1}{\Omega d_e}
   \langle \Psi_{\Omega}|\sum_iH_d(i)|\Psi_{\Omega} \rangle~,
\end{equation}
where $ \Psi_{\Omega} $ is the wavefunction for either $a(1)$ or $B(1)$, and
$\Omega= \langle\Psi_{\Omega}|\bm{J}\cdot\bm{n}|\Psi_{\Omega}\rangle$.
The hyperfine constant $A_{\parallel}$ is determined by the expression
\cite{Dmitriev:92}
\begin{eqnarray}
   A_{\parallel}=\frac{1}{\Omega} \frac{\mu_{\rm Pb}}{I}
   \langle
   \Psi_{\Omega}|\sum_i(\frac{\bm{\alpha}_i\times \bm{r}_i}{r_i^3})
_Z|\Psi_{\Omega}
   \rangle~,
 \label{All}
\end{eqnarray}
 where $\mu_{\rm Pb}$ is the magnetic moment of $^{207}$Pb, $\bm{\alpha}_i$
 are the Dirac matrices for the $i$th electron, and $\bm{r}_i$ is its
 radius-vector in a coordinate system centered on the Pb atom.

Both $A_{\parallel}$ and $W_d$ depend strongly on the electronic spin density
near the heavy nucleus, while the molecular bonds are formed in the valence
region.  As shown previously
(\cite{Titov:96b,Titov:96,Petrov:02} and
references therein), it is possible to evaluate the electronic wavefunction
near the heavy nucleus in two steps.  Using this strategy here, a
high-accuracy relativistic coupled cluster (RCC) calculation \cite{Kaldor:99}
of the molecular electronic structure with the generalized relativistic
effective core potential (GRECP) is carried out, providing proper
electronic density in the valence and outer core regions. This is followed by
restoration of the proper shape of the four-component molecular spinors in the
inner core region of the heavy atom.

%*************************************
\paragraph*{Methods and calculations.}

A 22-electron GRECP for Pb \cite{Mosyagin:97} is used in the first stage of
the two-step calculations of PbO: the inner shells of the Pb atom
($1s$ to $4f$) are absorbed into the GRECP, and the $5s5p5d6s6p$ electrons
and all the oxygen electrons are treated
explicitly.  Two series of calculations are carried out, denoted as
({\it a}) and ({\it b}):  calculation ({\it a}) correlates 10 electrons,
freezing the $5s5p5d$ shells of Pb and the $1s$ shell of O; ({\it b})
correlates all 30 electrons treated explicitly.
States with the leading configurations $\sigma_1^2\sigma_2^2 \pi_1^4$,
$\sigma_1^2\sigma_2^2 \pi_1^3 \pi_2^1$, and $\sigma_1^2\sigma_2^1 \pi_1^4
\pi_2^1$ are calculated. Here $\sigma_{1,2}$ and $\pi_{1,2}$ are molecular
valence orbitals, with the subscript enumerating them in order of increasing
energy.  For each series of calculations, correlation spin-orbital basis sets
are optimized in atomic two-component GRECP/RCC calculations of Pb.  The four
$6s$ and $6p$ electrons are correlated in the basis set optimization stage of
calculation ({\it a}), and 22 electrons ($5s$ to $6p$) are correlated in
the optimization of the basis set used in series ({\it b}).  Correlation
is taken into account at this stage by the RCC method with single and double
excitations (RCC-SD) \cite{Kaldor:97}; the average energy of the five
lowest states of Pb is minimized. The detailed description of the basis set
generation procedure may be found in Refs.\ \cite{Isaev:00,Mosyagin:00}. A
[$4s3p2d$] basis, obtained by omitting the $f$ function from Dunning's
correlation-consistent ($10s5p2d1f$)/[$4s3p2d1f$] basis listed in the
MOLCAS~4.1 library \cite{MOLCAS}, is used for oxygen. We found that the $f$
orbital has little effect on the core properties calculated here.
Previous calculations show that these basis sets are adequate for our purpose.

PbO calculations start with a one-component self consistent field (SCF)
computation of the molecular
ground state, using the spin-averaged GRECP (AGREP). The Pb spinors $5s5p5d$
are frozen in the ({\it a}) series, using the level-shift technique
\cite{Titov:99}. An AGREP/RASSCF (restricted active space SCF) calculation
\cite{Olsen:88,MOLCAS} of the lowest $^3\Sigma^+$ state of PbO is
then performed.  In the RASSCF method, orbitals are divided into three active
subspaces: RAS1, with a restricted number of holes allowed; RAS2, where all
possible occupations are included; and RAS3, with an upper limit on the number
of electrons.

Different distributions of electrons in these active subspaces are used
(details may be found in \cite{http}) to estimate the
different correlation contributions to the RASSCF values of $A_{\parallel}$ and
$W_d$. Two-component RCC-SD molecular calculations are then performed. The
AGREP/RASSCF calculations include only the most
important correlation and scalar-relativistic effects, while the GRECP/RCC-SD
calculations also account for spin-orbit interaction.
The Fock-space RCC calculations start from the ground state of PbO and use the
scheme
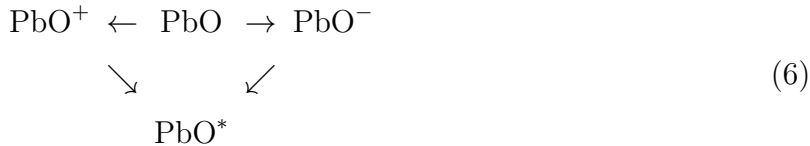
\begin{equation}
\begin{array}{ccccc}
 {\rm PbO}^+&\leftarrow&{\rm PbO}&\rightarrow&{\rm PbO}^-\\
              & \searrow&                 &\swarrow \\
              &         &{\rm PbO}^*&         \\
\end{array}
 \label{fockspacePbO}
\end{equation}
Details on the model spaces used may be found in
\cite{http}.

Only valence and outer core electrons have been treated so far.
Since we are interested in properties near the Pb nucleus,
the shape of the four-component molecular spinors
has to be restored in the inner core region. All molecular spinors are
restored using the nonvariational one-center restoration scheme (see
\cite{Titov:96b,Titov:99,Titov:00,Petrov:02} and references therein). This is
done in two steps:

 First, equivalent numerical one-center basis sets of four-component spinors and
 two-component pseudospinors are generated by the finite-difference
 all-electron Dirac-Hartree-Fock (DHF) and  GRECP/SCF calculations,
 respectively, of the same valence configurations of Pb and its
 ions.
 In the DHF calculations the inner core spinors (1s to 4f) are frozen
 after the calculation of Pb$^{2+}$, and
 the nucleus is modeled by a
 uniform charge distribution within a sphere of radius $r_{\rm nucl}=7.12{\rm
 fm}=1.35\times10^{-4}{\rm a.u}$.
 The root mean square radius of the nucleus is 5.52 fm, in accord with
 the parametrization of Johnson and Soff \cite{Johnson:85},
 and agrees with the $^{207}$Pb nucleus experimental value of 5.497 fm
 \cite{Fricke:95}. Taking the
 experimental value for the root mean square radius and a Fermi
 distribution for the nuclear charge changes $A_{\parallel}$
 and $W_d$ by 0.1$\%$ or less.
 The all-electron  four-component HFD \cite{Bratzev:77} and two-component
 GRECP/HFJ  \cite{Tupitsyn:95,Mosyagin:97} codes are employed for the basis
 generation, using the procedure developed in Refs.\
 \cite{Mosyagin:00,Isaev:00}. The basis sets generated are [$9s14p7d$] for
 series ({\it a}) and [$6s7p5d$] for series ({\it b}),
 with the latter carefully optimized.
 These sets are orthogonal to the inner core (see above). They
 describe mainly the core region, and are generated independently of
 the basis set for the molecular GRECP calculations discussed earlier.

 In the second step, the basis of one-center two-component atomic
 pseudospinors is used to expand the molecular pseudospinorbitals; these
 two-component pseudospinors are then replaced by the equivalent
 four-component spinors, retaining the expansion coefficients. A very good
 description of the wave function in the core region is obtained.

 The RCC-SD calculation of $W_d$ and $A_{\parallel}$ employs the finite field
 method \cite{Kunik:71,Monkhorst:77}.  The operator corresponding
 to the desired property [Eq.\ (\ref{eq:hd}) or (\ref{eq:hhf})] is multiplied
 by a parameter $\lambda$ and added to the Hamiltonian.  The first
 derivative of the calculated energy with respect to $\lambda$ gives
 the evaluated property.  This is strictly correct only at the limit of
 vanishing $\lambda$, but it is usually possible to find a range of
 $\lambda$ where the energy is linear in $\lambda$ and the energy
 changes are large enough to attain the required precision. The quadratic
 dependence of the energy on $\lambda$ is eliminated here
 by averaging the components of a given term, $a(1)$ or $B(1)$,
 with opposite signs of $\lambda$.

%***********************************
\paragraph*{Results and discussion.}

 Calculated results for the ({\it a}) and ({\it b}) series are presented in
 \tref{10e30e}. The internuclear distance is 2.0\,\AA.  The RASSCF
 calculations use the 22-electron GRECP for Pb. Twenty of the 30 electrons
 treated were in the inactive space, and only 10 were correlated. Using the
 C$_{\rm 2v}$ classification scheme, 2 A$_1$ orbitals are in RAS1, 6 orbitals
 (2 A$_1$, 2 B$_1$, and 2 B$_2$) in RAS2, and 41 (16 A$_1$, 5 A$_2$, 10 B$_1$,
 and 10 B$_2$) in RAS3. No more than two holes in RAS1 and two particles in
 RAS3 are allowed. The basis sets on Pb are ($14s18p16d8f$)/[$4s7p5d3f$] for
 the RASSCF and 30-electron RCC-SD calculations and
 ($15s16p12d9f$)/[$5s7p4d2f$] for 10-electron RCC-SD. A ($10s5p2d$)/[$4s3p2d$]
 basis is put on O in all calculations.

%\squeezetable
\begin{table*}
\caption
{
 Calculated  parameters $A_{\parallel}$ (in MHz) and $W_d$ (in
$10^{24}{\rm Hz}/(e\cdot {\rm cm})$)
for the $a(1)$ and $B(1)$ states of $^{207}$PbO.  The experimental value
of $A_{\parallel}$ in $a(1)$ is
$-4113 {\rm MHz}$. The preliminary value of $A_{\parallel}$
in $B(1)$ is $5000\pm 200 {\rm MHz}^{\rm 1}$
}
\begin{ruledtabular}
\begin{tabular}{lrrrrrrrrrrrrr}
State &\multicolumn{6}{c}{a(1)\ \ $\sigma_1^2\sigma_2^2 \pi_1^3 \pi_2^1$
\ \  $^3\Sigma_1$}&
&\multicolumn{6}{c}{B(1)\ \ $\sigma_1^2\sigma_2^1 \pi_1^4 \pi_2^1$\ \
$^3\Pi_1$}\\
Parameters &\multicolumn{3}{c}{$A_{\parallel}$} &&\multicolumn{2}{c}{$W_d$}
& & \multicolumn{3}{c}{$A_{\parallel}$}&& \multicolumn{2}{c}{$W_d$}\\
\cline{2-4}\cline{6-7}\cline{9-11}\cline{13-14}
Expansion & s& s,p& s,p,d&& s,p& s,p,d& & s& s,p& s,p,d&& s,p&
s,p,d \\
\hline

 10e-RASSCF
& -759& -1705 & -1699&& 0.96 & 0.91& & & &1900&&0.0 &0.0 \\

 10e-RCC-SD
& & & -2635&& -2.93& & & & &3878&& -11.1 & \\

 30e-RCC-SD
& -359& -3062& -3012&& -3.08& -3.18& & 195& 4510& 4568&& -10.4
&-9.7 \\

\end{tabular}
\end{ruledtabular}
$^{\rm 1}$ private communication, D.DeMille, 2003.\hspace{8.9cm}
\label{10e30e}
\end{table*}

 We discuss mainly the results for the $a(1)$ state (leading configuration
 $\sigma_1^2\sigma_2^2 \pi_1^3 \pi_2^1$), for which the 
 reliable experimental value of
 $A_{\parallel}$ is available ($-$4113~MHz) \cite{Hunter:02} and a
 semiempirical estimate of $|W_d|\ge 12\times 10^{24}$~Hz/($e\cdot$cm) was made
 recently \cite{Kozlov:02}. There are several points to note: (1) Inclusion
 of the spin-orbit interaction changes $A_{\parallel}$ and $W_d$ dramatically,
 as may be seen from the difference between the 10-electron RASSCF and RCC-SD
 results. (2) The {\it ab initio} value of $W_d$ is four times smaller than
 the semiempirical estimate \cite{Kozlov:02}. (3) Accounting for outer
 core--valence correlation by 30-electron RCC-SD changes $W_d$ by 5\% and
 $A_{\parallel}$ by 15\%, yet the error in the calculated $A_{\parallel}$ is
 25\%; calculations on BaF \cite{Kozlov:97} and YbF \cite{Mosyagin:98} gave
 10\% accuracy. (4) $A_{\parallel}$ is mainly determined by the $p$ wave,
 whereas $W_d$ mostly comes from $s$-$p$ mixing.

 The need for including correlation
 in the PbO molecule for the properties discussed here can be seen already  in
 the semiempirical model \cite{Kozlov:02}.
 The leading contribution to the highest occupied $\sigma_2$ orbital in this
 model comes from the Pb $6s$ atomic orbital, with a weight of $\sim$0.5 (the
 corresponding 
 coefficient in the molecular orbital expanded as a linear combination of
 atomic orbitals
 is $\sim$0.7). This contradicts
 the qualitative analysis of the chemical bond formation, which predicts that
 the $\sigma_2$ orbital is mainly formed from the oxygen $2p_\sigma$ and lead
 $6p_\sigma$ orbitals.  The RASSCF calculations of the lowest $^3\Sigma^+$
 state confirm this point, with the weight of the Pb $6s$ orbital varying
 between $0.04$ for 10 active electrons and $0.1$ for 30 active
 electrons. The weight of the oxygen $2p_\sigma$ is $\sim$0.5 and that of
 the lead $6p_\sigma$ is $\sim$0.1, whereas $\sigma_1$ consists mainly of
 the lead $6s$ orbital, with negligible contribution from lead $6p_\sigma$\,.
 Note that the oxygen $2p_\sigma$ and lead $6p_\sigma$ orbitals are not
 orthogonal to each other; after one-center reexpansion of the oxygen
 basis functions on lead (see~\cite{Titov:96} and Eq.~(6)
 in~\cite{Petrov:02}), the weight of the $6p_\sigma$ orbital goes up to 0.3.
 We expect that such strong admixture of the $s$-wave to the $\sigma_2$
 orbital would not appear in the semiempirical model if configurations
 describing the correlation of the $\sigma_2$ electrons were included in the
 model space.  It is important to add that the lowest virtual $\sigma_3$
 orbital gets the main contribution from the lead $6p_\sigma$ (with a weight
 of about 0.5), and the configurations containing this orbital are first
 admixed into the leading configuration of the $a(1)$ state due to the
 spin-orbit interaction on Pb.

 If the spin-orbit interaction is neglected, the $s$-wave
 contribution to $A_{\parallel}$ and the $s,p$-wave contributions to $W_d$ are
 due primarily to correlation of the $\sigma$ electrons.  The RASSCF
 calculation indicates (see \tref{10e30e}) that such contributions increase
 $A_{\parallel}$ but decrease $W_d$, resulting in a sign change for $W_d$, in
 agreement with the final RCC-SD result (details may be found in
 Ref.\ \cite{Petrov:03b}).  Besides, as correlation is expected to
 have a strong influence on the values of $A_{\parallel}$ and $W_d$,
 introducing the SO interaction with the $^3\Pi$ and $^1\Pi$ states by just
 mixing the corresponding $\sigma$ and $\pi$ orbitals may not be satisfactory.
 All these conclusions could be reached only after extensive molecular
 calculations, and the estimates made in Ref.~\cite{Kozlov:02} were important
 at the first stage of the experimental effort.

 As may be expected, the accuracy of the calculated $A_{\parallel}$ and $W_d$
 values is lower for such a complicated system as the excited states of the
 PbO molecule than for the ground states of BaF and YbF.  The valence electron
 in the latter molecules is in a $\sigma$ orbital, with much higher density
 near the heavy nucleus than the valence $\pi$ electrons in PbO.  Thus, the
 $s,p,d$-waves on the Pb nucleus are affected more strongly by correlation,
 and higher-order inclusion of correlation (triple and quadruple amplitudes in
 the RCC method) as well as larger basis sets may be necessary.  As pointed
 out above, $W_d$ is more stable than $A_{\parallel}$ with respect to changing
 the number of correlated electrons, and we expect the accuracy of the
 calculated $W_d$ to be better than for $A_{\parallel}$.  Our estimated error
 bounds put the real $W_d$ between 75\% and 150\% of the calculated value,
 which is quite satisfactory for the first stage of the EDM experiment on PbO.
 It should be noted that the estimate of statistical sensitivity to the
 electron EDM made in \cite{DeMille:00} is based on a $W_d$ value close to
 that obtained here.

 A detailed analysis of correlation and spin-orbital effects on
 $A_{\parallel}$ and $W_d$ in PbO will be published elsewhere
 \cite{Petrov:03b}.  Unfortunately, the experimentally available
 $A_{\parallel}$ of the $a(1)$ 
%Tim
 and $B(1)$ states
%state 
%end
 provides a check on the $p$ wave only.
 It would be desirable to measure $A_{\parallel}$ in some state with an
 excited $\sigma_1$-electron, with the main contribution coming from the
 $s$-wave.  Another accuracy check, using $\sqrt{A_{\parallel}A_{\perp}}$, is
 not applicable here, because of experimental difficulties in measuring the
 very small $A_{\perp}$ for diatomic molecules with total electronic momentum
 $J \geq 1$.  Our estimate of the accuracy of the calculated $W_d$ is
 therefore not as straightforward as for YbF and BaF \cite{Kozlov:97,
 Mosyagin:98}.

 Finally, we would like to note that  we identified the lowest $^3\Pi_1$ state
 as $B(1)$ according to the $\Lambda S$ classification given in
 \cite{Huber:79}. Conclusive identification requires more extensive {\it ab
 initio} correlation calculations.

%*****************************
\paragraph*{Acknowledgments.}

 The authors are very grateful to M.~Kozlov and D.~DeMille for many fruitful
 discussions and critical remarks.  Some codes of M.~Kozlov for the
 calculation of atomic properties were used in our PbO calculations as well.
 The present work is supported by RFBR grant No.\ 03-03-32335 and U.S.\ CRDF
 Grant No.\ RP2--2339--GA--02.  T.I.\ thanks INTAS for Grant YSF 2001/2-164.
 A.P.\ is grateful to the Ministry of Education of the Russian Federation
 (Grant PD02-1.3-236) and to the St.\ Petersburg Committee on Science and
 Higher Education (Grant PD03-1.3-60).  N.M.\ and A.T.\ are supported in part
 by the Scientific Program of St.-Petersburg Scientific Center of RAS.
 Research at TAU was supported by the Israel Science Foundation and the
 U.S.-Israel Binational Science Foundation.

\bibliographystyle{./bib/apsrev}
%\bibliography{bib/*}
\bibliography{bib/JournAbbr,bib/Titov,bib/TitovLib,bib/Kaldor,bib/Isaev}
\end{document}